\documentclass[10pt,twocolumn,reprint,amsmath,amssymb,amsfonts,aps,prd,superscriptaddress,showpacs,nofootinbib]{revtex4-2}

\usepackage{graphicx}
\usepackage{hyperref}
\usepackage[normalem]{ulem}
\usepackage{slashed}
\usepackage{float}
\usepackage[caption=false]{subfig}
\usepackage{bm}

\hypersetup{
	bookmarks=true,         
	unicode=true,          
	pdftoolbar=true,        
	pdfmenubar=true,        
	pdffitwindow=false,     
	pdfstartview={FitH},    
	pdftitle={GeVNPandgm2},    
	pdfauthor={Author},     
	pdfsubject={Subject},   
	pdfcreator={Creator},   
	pdfproducer={Producer}, 
	pdfkeywords={keyword1} {key2} {key3}, 
	pdfnewwindow=true,      
	colorlinks=true,       
	linkcolor=black,          
	citecolor=black,        
	filecolor=black,      
	urlcolor=black           
}

\usepackage[dvipsnames]{xcolor}

\usepackage{comment}
\usepackage[normalem]{ulem}

\newcommand{\eqn}[1]{Eq.~(\ref{#1})}

\newcommand{\eV}{\,\text{eV}}

\def\be{\begin{equation}}
\def\ee{\end{equation}}
\newcommand{\ba}{\begin{array}}
\newcommand{\ea}{\end{array}}
\def\bV{\begin{Verbatim}[breaklines=true, breakanywhere=true]}
\def\eV{\end{Verbatim}}

\def\amhvp{a_{\mu}^{\rm HVP}}
\def\amhvplo{a_{\mu}^{\rm LO, \rm HVP}}
\newcommand{\amc}{{\sc MadGraph5}\_a{\sc MC@NLO}}

\def\amIW{a^\text{HVP}_{\text{W}}}
\def\amLD{a^\text{HVP}_{\text{LD}}}
\def\amSD{a^\text{HVP}_{\text{SD}}}

\def\shad{\sigma_{\rm had}}

\newcommand{\HVPref}{Aoyama:2012wk,Aoyama:2019ryr,Czarnecki:2002nt,Gnendiger:2013pva,Davier:2017zfy,Keshavarzi:2018mgv,Colangelo:2018mtw,Hoferichter:2019mqg,Davier:2019can,Keshavarzi:2019abf,Kurz:2014wya,Melnikov:2003xd,Masjuan:2017tvw,Colangelo:2017fiz,Hoferichter:2018kwz,Gerardin:2019vio,Bijnens:2019ghy,Colangelo:2019uex,Blum:2019ugy,Colangelo:2014qya}

\allowdisplaybreaks[1]

\newcommand*{\INFNFR}{Istituto Nazionale di Fisica Nucleare, Laboratori Nazionali di Frascati, C.P. 13, 00044 Frascati, Italy}
\newcommand*{\IPII}{Universit\'e de Lyon, Universit\'e Claude Bernard Lyon 1, CNRS/IN2P3,\\
Institut de Physique des 2 Infinis de Lyon, UMR 5822, F-69622, Villeurbanne, France}
\newcommand*{\INFNRM}{Istituto Nazionale di Fisica Nucleare, Sezione di Roma, Piazzale A. Moro 2, I-00185 Roma, Italy}
\newcommand*{\TALLIN}{ Laboratory of High Energy and Computational Physics, HEPC-NICPB,  R\"avala 10, 10143 Tallinn, Estonia}

\begin{document}

\title{Indirect new physics effects on $\shad$ confront the (g-2)$_\mu$  window discrepancies and the CMD-3 result}

\author{Luc Darm\'e}\email{l.darme@ip2i.in2p3.fr}\affiliation{\IPII}
\author{Giovanni Grilli di Cortona}\email{grillidc@lnf.infn.it}\affiliation{\INFNFR}\affiliation{\INFNRM}	
\author{Enrico  Nardi}\email{enrico.nardi@lnf.infn.it}\affiliation{\INFNFR}	\affiliation{\TALLIN}

\begin{abstract}
Recent lattice determinations of the hadronic vacuum polarization contribution to the muon  anomalous magnetic moment  $\amhvp$ have confirmed the discrepancy with the data-driven dispersive method. In the meanwhile the  CMD-3 collaboration has reported a result for the $e^+e^-\to \pi^+\pi^-$ cross section considerably larger than previous experimental results (and close to the lattice determinations) exacerbating the discordance between different $e^+e^-$ datasets. We explore to what extent these disagreements can be accounted for by some new physics effect altering selectively the individual experimental determinations of $\sigma(e^+e^- \to\;$hadrons).
We find that specific effects of GeV-scale new particles are able to shift upwards the KLOE and BaBar results in the low and intermediate energy windows, while leaving unaffected the CMD-3 energy scan. Although these new physics effects cannot fully explain all the discrepancies among the different $\sigma(e^+e^- \to\;$hadrons) datasets, they succeed in mitigating the overall tension between data-driven and lattice  estimates of $\amhvp$. Remarkably, the additional loop corrections involving  the new particles concur to solve the residual discrepancy with the experimental value of $(g-2)_\mu$.
 \end{abstract}

\maketitle

\tableofcontents


\section{Introduction}

The theoretical uncertainty in the Standard Model (SM) prediction for the anomalous magnetic moment of the muon $a_\mu$  is currently dominated by the error associated with the 
hadronic vacuum polarization (HVP)  contribution $\amhvp$. 
The recommended value 
for the leading order HVP correction~\cite{Aoyama:2020ynm}  is based on $e^+e^- \to$ hadrons cross section data, 
and reads:
\be
\label{eq:datadriven}
\amhvplo\biggl|_{\mathrm{data-driven}}=693.1(4.0)\cdot 10^{-10}.
\ee
This yields the SM prediction~\cite{\HVPref}:
\be
a_\mu^{\textrm{SM}}=11659181.0(4.3)\cdot 10^{-10}.
\ee
Comparing this result  with the current experimental world average obtained by combining the previous BNL~\cite{Bennett:2006fi} and FNAL\cite{Abi:2021gix} results with 
the new FNAL determination~\cite{Muong-2:2023cdq}:
\be
\label{eq:amuexp}
a_\mu^{\textrm{exp}}=116592059(22)\cdot 10^{-11},
\ee
leads to a tension at the  level of 5$\sigma$.

On the other hand, $\amhvp$   can be also computed from first principles 
by means of QCD lattice techniques. 
The most precise lattice result, obtained  by the BMW collaboration~\cite{Borsanyi:2020mff}, is 
\be
\amhvplo\biggl|_{\mathrm{BMW}}=707.5(5.5)\cdot 10^{-10}\,
\ee
which is in tension at 2.1$\sigma$ with the data-driven determination in \eqn{eq:datadriven} and, most noticeably, 
would reduce the difference with  the experimental value in \eqn{eq:amuexp} to 1.5$\sigma$.

Clearly, a confirmation of the correctness  of the BMW result would have a major impact on 
assessing the need for new physics (NP) to account for  the measured value of $a_\mu$. 
Unfortunately,  lattice-QCD results are generally affected by large systematic and statistical 
uncertainties mostly related to the infinite volume and continuum limits, which 
have so far prevented high accuracy determinations of  the HVP. 
However,   specific  parameter space regions exist in  which the previous uncertainties  
are under better control, and within these regions more precise  results can 
be obtained.  
The so-called short,  intermediate and long  Euclidean  time-distance windows (respectively labelled with SD, W and LD)
 were first defined by the RBC/UKQCD collaboration in Ref.~\cite{RBC:2018dos}. Weight functions are introduced in  
the HVP integrals, which  allow  to select only certain regions in parameter space, and in particular  regions that are less affected by the sources of uncertainty. 
In recent years, precise  lattice-QCD determination of 
partial HVP integrals  became available.
The determination of the HVP contribution of the intermediate 
window $\amIW$  provided  by the 
BMW collaboration \cite{Borsanyi:2020mff},   
by the CLS/Mainz group \cite{Ce:2022kxy}, 
by the Extended Twisted Mass Collaboration (ETMC) \cite{Alexandrou:2022amy}, 
 by Lehner and Meyer \cite{Lehner:2020crt},
  by Aubin et al.  \cite{Aubin:2022hgm} 
 (which updates their previous result  \cite{Aubin:2019usy}),
by the $\chi$QCD collaboration \cite{Wang:2022lkq} and, 
very recently,
by the Fermilab Lattice, HPQCD, MILC~\cite{Bazavov:2023has}
and  RBC/UKQCD Lattice Collaborations~\cite{Blum:2023qou}, 
are in overall good agreement, giving strong support to the reliability of lattice evaluations.
At the same time they shed some motivated suspicion on the result of the 
dispersive method  which, in the same window \cite{Colangelo:2022vok}, 
is several $\sigma$ below the average of the lattice results.\footnote{The tension 
between the dispersive method~\cite{Colangelo:2022vok} and  the individual results 
of  different lattice collaborations is  around   $4\sigma$~\cite{Borsanyi:2020mff,Ce:2022kxy,Alexandrou:2022amy,Blum:2023qou}.  
Ref.~\cite{Alexandrou:2022amy}  quotes a $4.5\sigma$ tension
for the combined BMW,  CLS/Mainz, and  ETMC  results neglecting correlations. 
Ref.~\cite{WittingMoriond2023} quotes a $3.8\sigma$ tension for the 
 combined BMW,  CLS/Mainz,   ETMC and RBC/UKQCD assuming 100\% correlation.}
In contrast,  in the short distance window no substantial discrepancy has been encountered~\cite{Alexandrou:2022amy} (see also~\cite{FermilabLattice:2022izv}). 
The deviation between the lattice results and the data-driven determination may thus  
  be interpreted as an   effect of NP that is localised in 
  the intermediate and possibly large distance windows. 

In Ref.~\cite{Darme:2021huc} it was pointed out that, besides  {\it direct} NP loop 
contribution to  $(g-2)_\mu$, 
additional NP effects  acting {\it indirectly} on the way 
$\shad$ is extracted from the experimental data were needed in order to reconcile 
the various discrepancies known at that 
time,\footnote{The need to resort to {\it indirect} effects on the measurement of $\shad$
to account for the $a_\mu$ discrepancies is also warranted by the fact that,
as was argued in Ref.~\cite{DiLuzio:2021uty} (see also Ref.~\cite{Crivellin:2022gfu}), 
the possibility of solving the lattice/data-driven discrepancy relying only on 
{\it direct} NP  contributions  to  the $e^+ e^- \to\,$hadrons process 
is excluded by a number of experimental constraints,
as for example the global electroweak fits, which would enter in serious 
tension with observations, because of  modifications  
of the  hadronic contribution to the running of the fine-structure 
constant~\cite{Passera:2008jk,Crivellin:2020zul,Keshavarzi:2020bfy,Malaescu:2020zuc}.}
and a class of NP scenarios in which specific types of indirect effects can affect the various experiments 
in different ways was put forth.

This is an important feature  in view of the fact that 
 besides the 4.2$\sigma$ discrepancy between the data-driven SM prediction for $ a_{\mu}$ 
and the experimental result and the  discrepancy of comparable significance 
with the lattice results for $\amIW$, significant disagreements between different experimental 
determinations of $\shad$ are also present.
In particular, the long-standing $\sim\!3\sigma$ discrepancy between KLOE and BaBar, which  provided 
the two most accurate determinations of $\shad$,
is now overshadowed by the new CMD-3 measurement of the $e^+e^- \to \pi^+\pi^-$ cross 
section~\cite{CMD-3:2023alj}.\footnote{It should be remarked that a recent study  of 
higher-order radiative processes in $e^+e^-\to \mu^+\mu^-\gamma$ and $e^+e^-\to\pi^+\pi^-\gamma$   
events  performed by the BaBar collaboration  at NNLO~\cite{BaBar:2023xiy}, pointed out that 
the different treatment of these events adopted by different experiments that 
determine $\shad(s)$ thorough the radiative method  may explain the  discrepancy
between KLOE and BaBar. However, the discrepancy with CMD-3 that is using the scanning method 
cannot be explained in a similar way. \label{foot:BaBar}} 
The result of this measurement yields a value of $\sigma_{\pi^+\pi^-}$ well above previous results.
In particular, in the energy range $\sqrt{s} \in [0.6,0.88]\,$GeV the CMD-3 contribution to $\amhvplo$ is 
more than $5\,\sigma$ above  
the contribution estimated from KLOE data~\cite{Anastasi:2017eio,CMD-3:2023alj}. 

The aim of this work is to study in details whether the class of new physics scenarios introduced in  Ref.~\cite{Darme:2021huc} can:
\begin{enumerate}
\item solve the discrepancy between the data-driven and lattice evaluation of the HVP contribution $\amIW$ in the intermediate window;
\item solve the discrepancy between the full experimental measurement $a_\mu^{\textrm{exp}}$ and its data-driven counterpart;
\item improve the consistency between the different data-sets used for the the data-driven estimate of the HVP contribution $\amhvplo$. 
\end{enumerate}

We will thus study the impact of indirect NP effects on the intermediate 
energy window $\amIW$ for the different data-sets that we  label as 
KLOE08~\cite{KLOE:2008fmq}, KLOE10~\cite{KLOE:2010qei} 
 KLOE12~\cite{KLOE:2012anl},  BESIII~\cite{BESIII:2015equ}, 
 BaBar~\cite{BaBar:2009wpw,BaBar:2012bdw} (for which we perform for the first time a full study of the published results)
 and CMD-3 (for which we estimate the $\amIW$ contribution 
 using the data in the ancillary files of Ref.~\cite{CMD-3:2023alj}). 

Finally, we also include in the final fit the most 
 recent SND 2020 result~\cite{SND:2020nwa}, the results from 
 CMD-2~\cite{Akhmetshin:2003zn,Akhmetshin:2006wh,Akhmetshin:2006bx} and the older SND 2006 measurement~\cite{Achasov:2006vp}.

This work extends the study of Ref.~\cite{Darme:2021huc} in several important ways.  First, we include a simulation of the BaBar analyses in the \amc\ platform~\cite{Alwall:2014hca} that uses the muon method for the determination of $\shad$~\cite{BaBar:2009wpw,BaBar:2012bdw}.\footnote{The BaBar analysis requires a \textit{visible} photon in the detector acceptance and two reconstructed muons tracks. The corresponding NP process is thus $e^+e^- \to \gamma V$, along with the semi-visible $V$ decay. More details  about the reconstruction procedure used for BaBar are given in Appendix~\ref{sec:app1}.} Secondly, we estimate the contribution to $ \amhvp$ in the intermediate 
energy window from the  CMD-3 measurement reported in Ref.~\cite{CMD-3:2023alj}, and we compare this new piece of information with the corresponding results from other experiments and from the lattice. Third, we refine the simulation of the experimental efficiencies of the NP signal. 

All in all, we find that within our scenario a good consistency between  $(g-2)_\mu$ values inferred from lattice, data-driven, and experimental results can be recovered, although the consistency between the different $\shad$ data-sets can only be marginally improved.
In particular, we find that the dominant repercussions from NP processes 
on the determination  of $\shad$  are confined to the 
low and intermediate energy windows 
while the high energy window remains largely unaffected,
which is in nice agreement with lattice indications. 
A final remark is in order regarding the recent  CMD-3 measurement of $\sigma_{\pi^+\pi^-}$. Since this measurement is performed with the energy-scanning method in the energy range $\sqrt{s} \in [0.6,0.88]\,$GeV~\cite{CMD-3:2023alj}, that is 
at CoM energies well below the $V$ resonance, it is clear that 
their result cannot be affected by the NP construction of Ref.~\cite{Darme:2021huc}.
Hence, in our scenario the good agreement between CMD-3 and the lattice results, 
and the marked disagreement with other experiments performed 
with the radiative method at  $\sqrt{s} \sim M_V $ or at $\sqrt{s} \gg M_V$, has a natural explanation.

In Sec.~\ref{sec:window_descr} we briefly summarise the experimental and lattice status and introduce the time 
windows' kernels. In Sec.~\ref{sec:NP} we discuss the various indirect effect that GeV-scale NP can have on 
the determination of $\shad$ with the
 dispersive approach. 
 In Sec.~\ref{sec:model} we introduce a
 phenomenological NP scenario  
 wherein a viable solution to all the $a_\mu$ window discrepancies
 can be provided.
 Finally, in Sec.~\ref{sec:conclusions} we draw our conclusions.

 \begin{table}[t]
\renewcommand{\arraystretch}{1.2}
	\centering
	\scalebox{0.9}{
	\begin{tabular}{l | c | c | c | c}
	 & $\amSD$ & $\amIW$ & $\amLD$ & $\amhvp
	 $\\
	 \hline
	 \hline
	Data-driven~\cite{Colangelo:2022vok} & $68.4(5)$ & $229.4(1.4)$ & $395.1(2.4)$ & $693.0(3.9)$\\
	BMWc~\cite{Borsanyi:2020mff} & -- & $236.7(1.4)$ & -- & $707.5(5.5)$\\
	Mainz/CLS~\cite{Ce:2022kxy} & -- & \!$237.30(1.46)$ & -- & --\\
	ETMC~\cite{Alexandrou:2022amy} & $69.27 (34)$ & $236.3 (1.3) $ & -- & --\\
 RBC/UKQCD~\cite{Blum:2023qou}  & -- & \!$235.56 (82) 
 $ & -- & --\\
 	Lattice average~\cite{WittingMoriond2023} & -- & $236.16 (1.09)$ & -- & --\\
	\hline
	\end{tabular}
	}
	\caption{
 Results for the short distance, intermediate and long distance   windows contributions and for the total $\amhvp$
  contribution to $a_\mu$. The results for the data-driven approach~\cite{Colangelo:2022vok} are given in the first line, and QCD lattice 
  results~\cite{Borsanyi:2020mff,Ce:2022kxy,Alexandrou:2022amy,Blum:2023qou} in the following lines. The lattice average from Ref.~\cite{WittingMoriond2023} 
  assumes 100\% correlation. 
	All numbers are in units of $10^{-10}$.}
	\label{tab:Wsummary}
\end{table}

\section{Time windows' kernels and the data-driven approach}
\label{sec:window_descr}

The full HVP contribution to $a_\mu$ can be decomposed as the sum of three terms corresponding to the three windows SD, W and LD~\cite{RBC:2018dos}:

\be
\amhvp \equiv  \amSD+ \amIW + \amLD
\ee
that on the 
QCD lattice 
correspond to different Euclidean time windows. Each term is obtained by modifying the integration kernel using predefined smooth step-functions in order to exponentially suppress contributions from other regions.
In the time-momentum representation, the HVP contribution to the muon anomalous magnetic moment is
\be
\amhvp = \left( \frac{\alpha}{\pi}\right)^2 \int_0^\infty \! \! dt \, \tilde{K}(t) G(t),
\ee
where $\tilde{K}(t)$ is a kernel function (given in Appendix B of Ref.~\cite{DellaMorte:2017dyu}) and $G(t)$ is given by the correlator of two electromagnetic currents. The windows in Euclidean time 
are defined by means of an  
additional weight function:
\begin{align}
\Theta_{\rm SD}(t) &= 1-\Theta(t,t_0,\Delta),\nonumber \\
\Theta_{\rm W}(t) &= \Theta(t,t_0,\Delta) - \Theta(t,t_1,\Delta),\nonumber \\
\Theta_{\rm LD}(t) &= \Theta(t,t_1,\Delta),\nonumber \\
\Theta(t,t',\Delta) &= \frac{1}{2}\left( 1 + \tanh\frac{t-t'}{\Delta} \right)\, ,
\end{align}
with parameters $t_0=0.4$ fm, $t_1=1.0$ fm and $\Delta=0.15$ fm.
In order to compare with the data-driven approach, the HVP contributions can be written as
\be
a_i^{\mathrm{HVP}} = \frac{1}{2\pi^3} \int_{E_{\mathrm{th}}} E\, \hat{K}\left(\frac{E}{m_\mu}\right)\, \sigma_{\mathrm{had}}(E) \,\hat{\Theta}_i(E) \,dE,
\ee
where $E$ is the $e^+e^-$ CoM energy, $E_\text{th}$ the threshold energy, 
$\hat{K}(x)=\int_0^1 dy \frac{(1-y)\,y^2}{y^2+(1-y)x^2}$ is the kernel function,   
$m_\mu$ the muon mass, $\sigma_\text{had}=\sigma(e^+e^- \to \text{hadrons})$, 
the index $i=$  LD,\,W,\,SD refers to a specific window,  
and~\cite{Colangelo:2022vok,Alexandrou:2022amy}
\be
\hat{\Theta}_i(E)=\frac{\int_0^\infty dt\,t^2\,e^{-E\,t}\,K(m_\mu\,t)\,\Theta_i(t)}{\int_0^\infty dt\,t^2\,e^{-E\,t}\,K(m_\mu\,t)}, 
\ee
where the kernel function is defined as 
$$
K(z)=2\int_0^1\,dy(1-y)\left[1-j_0^2\left(\frac{z\,y}{2\sqrt{1-y}}\right)\right],
$$ 
with $j_0(x)=\sin(x)/x$.

\begin{figure}[t!]
\vspace{-10pt}
    \centering
    \includegraphics[width=0.48\textwidth]{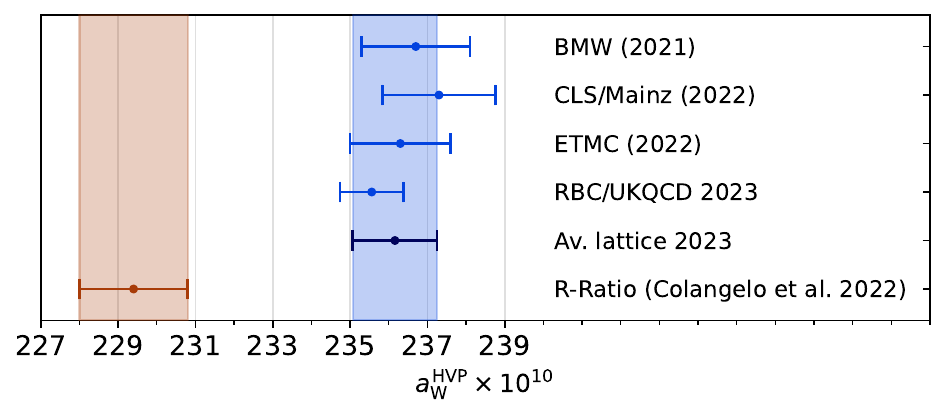}
    \caption{Comparison of the results for the intermediate window contribution to $\amhvp$ 
   from  the data-driven approach~\cite{Colangelo:2022vok} (region in red) and   
         from four lattice computations (BMW ~\cite{Borsanyi:2020mff},  CLS/Mainz~\cite{Ce:2022kxy} and ETMC~\cite{Alexandrou:2022amy}, RBC/UKQCD~\cite{Blum:2023qou} in blue). The black point and the blue region correspond to the average  given in Ref.~\cite{WittingMoriond2023}  of the 
      four lattice results assuming conservatively a 
          100\%          correlation.} 
             \label{fig:summary_IW}
\end{figure}

In Table~\ref{tab:Wsummary}, we collect the results for the windows contributions to  the 
HVP taken  from Refs.~\cite{Colangelo:2022vok,Borsanyi:2020mff,Ce:2022kxy,Alexandrou:2022amy,Blum:2023qou}. 
We see that while there are no indications  of sizeable discrepancies in the short distance window, in the  intermediate window the disagreement between the lattice  
and the data-driven results is remarkable.
Figure~\ref{fig:summary_IW} summarises  the situation  for the  intermediate window. 
The data-driven result of~\cite{Colangelo:2022vok} corresponds to the red data point and shaded region, the four  lattice computations~\cite{Borsanyi:2020mff,Ce:2022kxy,Alexandrou:2022amy,Blum:2023qou} correspond to the black data points, and their average~\cite{WittingMoriond2023} obtained by 
 assuming conservatively a 
          100\%          correlation
correspond to the blue data point and shaded region. It is clear that 
the $2.1\,\sigma$ discrepancy between the  data-driven and the BMW lattice 
evaluations of the total $\amhvp$ is exacerbated in the intermediate window. Most importantly, the BMW result for $\amIW$ is confirmed independently by the recent results of the Mainz/CLS~\cite{Ce:2022kxy},   ETMC~\cite{Alexandrou:2022amy},   
RBC/UKQCD~\cite{Blum:2023qou} and 
Fermilab, HPQCD, and MILC~\cite{Bazavov:2023has} lattice collaborations, strengthening the confidence in the reliability of the lattice approach.

\section{GeV-scale new physics indirect effects on the data-driven determination}
\label{sec:NP}

The data-driven method relies on the assumption that all the processes that concur to 
determine $\shad$ are SM processes. 
In presence of GeV-scale new physics, and in particular in the case 
the relevant  couplings are sizeable enough to affect the muon $(g-2)_\mu$, this hypothesis can fail at multiple levels.

To give an example, when the hadronic cross section is directly inferred from the 
detected number of hadronic events, a relation similar to the following one is generally used: 
\begin{equation}
    \label{eq:shad}
     \frac{d \shad}{d s^\prime}  = \frac{N_{\rm had} -
      N_{\textrm{bkd}}}{\epsilon ( s^\prime)\; \mathcal{L} ( s^\prime)}\;,
\end{equation}
where $N_{\rm had}$  
is the   measured number of hadronic final states produced in $e^+e^-$ annihilation with final-state invariant mass $\sqrt{ s^\prime}$, 
$N_{\textrm{bkd}}$ is the estimated number of background events,
$\epsilon ( s^\prime)$ is the   detection efficiency,  and  $\mathcal{L}(s^\prime)$ the 
luminosity for final states with invariant mass $\sqrt{ s^\prime}$, 
corresponding to the CoM energy of the collision (which,
in experiments that exploit initial state radiated photons to scan over $s'$,
is   different from the electron/positron CoM beam energy $\sqrt{s}$). 
A crucial observation is  that  each one of these  quantities can be affected by the presence of NP. In particular   $\mathcal{L}(s^\prime)$  is  generally determined by comparing the measurements of some other process
(e.g. Bhabha scattering) with the SM theoretical expectation.

\subsection{Luminosity estimate from Bhabha scattering}
In order to extract the 
hadronic cross-section $\shad$ at the sub-percent level, the experimental luminosity must be precisely measured. 

In 
the two earlier analysis of the KLOE collaboration~\cite{KLOE:2008fmq,KLOE:2010qei} (referred to as KLOE08 and KLOE10),   
the luminosity was inferred by comparing  measurements of the  Bhabha cross section  $e^+ e^- \to e^+ e^-$ at large angles  
with the SM predictions from  high precision Bhabha event generators~\cite{Jadach:1996gu,Arbuzov:2005pt,Balossini:2006sd}. 
As discussed extensively in~\cite{Darme:2021huc},  a  new physics contribution  to  Bhabha scattering 
able to affect the determination of $\shad$  at the required level via an incorrect determination of 
$\mathcal{L} ( s^\prime)$, should be  
of the order of  $\sim \mathcal{O} (1\, \textrm{nb})$. This  can be obtained for instance by resonantly producing 
a new  boson  of mass close to the KLOE8/KLOE10  CoM energies.  
In the following, we will assume  
a phenomenological setup in which a  dark photon (DP) $V$ with a mass around 
the GeV,  decays semi-visibly yielding  the required  excess of $e^+e^-$ events,
together with some missing/energy momentum associated with additional 
invisible decay products.

A remark is in order regarding the 
present treatment of the BESIII dataset 
with respect to our previous analysis~\cite{Darme:2021huc}.
The most recent measurement of BESIII~\cite{BESIII:2015equ} relies on Bhabha scattering to calibrate the luminosity instead of the ratio with muon final states (both methods are used in the paper, but the Bhabha approach is eventually preferred due to the smaller experimental errors). Thus,  the overall shift that we obtain   for BESIII in the present analysis  is sizeably  smaller than in our previous analysis. However, this does not impact strongly the overall fit due to the relatively large error bars of the BESIII~\cite{BESIII:2015equ} measurement compared to the KLOE and BaBar results. 

\subsection{The \texorpdfstring{$\sigma(\mu\mu\gamma)$}{smumug} method}
More recent measurements, including KLOE12~\cite{KLOE:2012anl} and BaBar~\cite{BaBar:2012bdw} 
estimate  the luminosity directly from the number of di-muon final states 
which can be collected  along with a $\pi \pi \gamma$ dataset. 
While this approach allows for the approximate cancellation of many systematic uncertainties,  
 it relies much more critically on the SM-only hypothesis due the much smaller SM $\mu\mu\gamma$ cross-section compared to Bhabha.  
 NP  effects do not require a tuning of the masses 
of the new particles involved  to be relevant for affecting analyses that adopt  this strategy. Relevant  effects  can be quite generically expected for any GeV-scale new boson with 
a coupling to  muons sufficiently large  to explain the  $(g-2)_\mu$ anomaly via new loop contributions. 
Since $\shad$ is eventually obtained by multiplying the ratio of events 
$N_{\pi\pi\gamma}/N_{\mu\mu\gamma}$ by the  theoretical SM muon production cross section
$\sigma_{\mu\mu}^{\rm SM}$, 
any excess of NP-related $\mu\mu X$ events must be subtracted from the 
dataset in order to obtain 
the correct value of $\shad$. 
If, under the assumption of SM $\mu$-production only, 
this is not done,  the inferred value of $\shad$ will be lower, 
given that the NP-related $\mu^+\mu^-$ events contribute to the normalization 
factor of the  hadronic events.  
If instead the NP origin of some $\mu^+\mu^-$ events is 
accounted for, the net effect is to increase the inferred value of $\shad$.

In a general DP model, decays of the hypothetical boson, besides additional $\mu^+\mu^-$
events,  
may also lead to additional hadronic final states. However, the fact that at the $\rho/\omega$ peak 
 the SM-rate for the hadronic processes are larger than that for $\mu^+\mu^-$  production 
 by an order of magnitude, implies that the NP effects in the hadronic channel are much less important than in the muon channel.  
To clarify further this point, let us consider two-pion events of NP origin 
 that, because of the missing energy/momentum associated 
 with the  $V$ invisible decay products, are reconstructed 
 at invariant masses laying  within the $\rho$ region. 
These events are produced via excitation of the $V$ resonance at $\sqrt{s} \sim 1\,$GeV and as a result, their rate is not increased by the hadronic resonances that  require $\sqrt{s'} \sim m_\rho$ to contribute to enhance the 
two-pion channel.\footnote{In fact,  the attempt  to fit the anomaly via the {\it direct} interference of NP contributions 
to the $e^+ e^- \to\,$hadrons process
carried out 
in Ref.~\cite{DiLuzio:2021uty} 
(see also Refs.~\cite{Crivellin:2022gfu,Coyle:2023nmi}) requires (using our notations) 
$|\epsilon_e (\epsilon_u-\epsilon_d)| \sim 2 \cdot 10^{-2}$, that is couplings  that are much  larger than those required to render effective our indirect effects. }

If we assume universality for the $V$ couplings 
(for example proportionality to the electromagnetic charge), than the number of muonic  and hadronic final states of NP origin that are reconstructed with invariant mass within the $\rho$ region 
will be roughly comparable. As a result, this effect will be much more significant for the $\mu$ 
channel than for  the $\pi$ channel. 
The dominant effect of  $V$-quark interactions is in fact 
that of reducing the branching ratio for decays yielding  
 di-muons in the final state. 
A naive estimate of this effect by accounting for  hadronic events induced by off-shell $V^*$ mixing with the $\rho$ 
leads  to a reduction of the NP shift by roughly one half. 
However, an additional complication is that if NP contributes to the hadronic channel, 
then the inferred $\shad = \shad^{QED} + \shad^{NP}$ cannot be directly 
related to the photon HVP, since one should first extract the pure QED contribution. 
While keeping in mind these caveats, in Section~\ref{sec:model} we will simply 
adopt a phenomenological model in which it is assumed that the $V$ coupling to the 
pions are sufficiently suppressed with respect to the couplings to the charged leptons, leaving for future work a detailed  estimate of hadronic effects in the representative case 
in which $V$-quark couplings are fixed by some 
universality condition.

\subsection{Background subtraction} 
The precise subtraction of background events is a key issue in most experimental analyses, since only the \emph{true} hadronic final states must be retained, while  
spurious events must be identified and rejected.
For instance, an important background considered by KLOE for the analysis using the $\mu \mu \gamma$ events  is the $\pi^- \pi^+ \pi^0$ final state,  as well as $\mu \mu \gamma$ events that 
can also be mis-identified as $\pi^- \pi^+ \gamma$.
This issue may be particularly important for the KLOE analysis, due to the fact that the particle identification between muons and pions relied on the so-called 
\emph{computed track mass} $m_{tr}$. The latter is defined  in terms of the  momenta $p_+$ and $p_-$ of the reconstructed positively and negatively charged tracks, based on the energy conservation relation
\begin{align}
\label{eq:mtr}
    \left(\sqrt{s}-\sqrt{|\vec{p}_+|^2+m^2_{tr}}-\sqrt{|\vec{p}_-|^2+m^2_{tr}} \right)^2 -|\vec{p}_- + \vec{p}_+|^2 = 0 \, ,
\end{align}
which assumes a SM process containing  a real photon with $E_\gamma = |\vec p_\gamma| = -|\vec{p}_- + \vec{p}_+|$. 
For NP  events with additional missing energy, as  for example the four body process $e^+e^-\to V \to \mu^+\mu^- \chi_1 \chi_1$ 
where $V$ is a hypothetical DP produced on shell whose decay products also contain the invisible 
particle $\chi_1$, 
energy conservation would imply the replacement 
$|\vec{p}_- + \vec{p}_+|^2 \to (E_{\chi}+E_{\chi'})^2 $. Experimentally,  muons and pions from these events would thus tend to yield   track mass solutions with values  somewhat larger than for the SM process.  
In fact, a full simulation shows that the track mass distribution for our $\mu \mu \chi_1 \chi_1$ final states are roughly flat, with a lower threshold at the muon mass. As the KLOE collaboration eventually rely on a fit on Monte Carlo (MC)-based distributions to distinguish the $\mu \mu \gamma, \pi \pi \gamma$ and $\pi^+ \pi^- \pi^0$ sample, the effect of injecting a NP signal which does not directly match any of these distributions cannot be easily estimated, in the lack of access to the  simulation tools used by the collaboration.

\subsection{Efficiencies} 
All analyses rely to a certain extent to a tag-and-probe approach to derive their efficiencies from the data. This two step process works as follows:
\begin{itemize}
    \item The selection cuts are applied on the dataset, requiring only one $\mu/\pi$ track. When available, particle identification requirements are made more stringent on this track. This forms the \textit{data control sample}.
    \item A kinematic fit is performed on the track along with the reconstructed photon to determine the most likely localisation for the opposite charge $\mu / \pi$, with the tagging and reconstruction efficiencies obtained by comparing with the reconstructed event.
\end{itemize}
Critically, the differences between the MC and the data on this sample are eventually used to apply a mass-dependent data/MC correction. If NP events are included in the data control sample, the efficiency estimate will be biased. Additionally, since the NP does not necessarily treat $\pi \pi$ and $\mu \mu$ final states on equal footing, there is no reason to expect that the efficiencies will cancel in the ratios between $\pi\pi$ and $\mu\mu$ events, as is broadly expected in the SM.
In the KLOE12 analysis, these efficiencies have been found to agree with the MC simulation within a few per mil, leaving little room for a significant NP  effect. On the other hand, corrections at a few percents are used in the BaBar analysis. A complete study of the potential NP effects in the determination of the efficiencies should be undertaken directly by the experimental collaborations.

\section{Explicit model and results}
\label{sec:model}

An example of how the NP contributions to Bhabha and $e^+e^-\to \mu^+\mu^-$ events 
needed to account for the various $a_\mu$-related discrepancies (while evading all other experimental constraints) was provided in Ref.~\cite{Darme:2021huc}, 
that adopted the inelastic dark matter model of Refs.~\cite{Izaguirre:2015zva,Darme:2017glc,Berlin:2018jbm}
in which a dark Abelian gauge group $U(1)_D$, kinetically mixed  with $U(1)_{\rm QED}$, is spontaneously broken by the vacuum expectation value of a dark Higgs $S$.
For simplicity here we assume a simple phenomenological setup   
along the lines of the model of Ref.~\cite{Feng:2016ysn} 
in which a new $J^P=1^-$  vector $V$  couples to a 
current $J^\mu_V$ that is a linear combination of the
SM fermion currents. In addition, we assume that $V$ 
has also an off-diagonal coupling to two Majorana dark fermions  $\chi_{1,2}$:
\begin{equation}
\mathcal{L} \supset - e V_\mu J^\mu_V 
- g_D V_\mu \,\bar\chi_2 \gamma^\mu \chi_1\,, \quad
J^\mu_V =  \sum_{i=u,d,\ell,\nu}  \epsilon_i \bar f_i \gamma^\mu f_i \,, 
\label{ea:lag}
\end{equation}
where $e$ is the usual QED coupling while $g_D$ is the $V$ coupling to the 
dark sector fermions. 
For the parameters appearing in $J_\mu^V$ we assume $\epsilon_i \ll 1 $ and, in particular,  
a  relative suppression of the neutrino with respect to the charged leptons couplings,  
 sufficient to evade the constraints from neutrino trident production~\cite{Altmannshofer:2014pba}
as well as other neutrino-related  constraints~\cite{Bauer:2018onh}, 
that is $\epsilon_\nu < \epsilon_\mu \simeq \epsilon_e \equiv \varepsilon$
(where the parameter  $\varepsilon$ should not be confused with   $\epsilon$ that refers 
instead to the efficiency). 
We also assume a certain suppression of the $V$ couplings to the light quarks
($\epsilon_{u,d} \lesssim \varepsilon$) 
to justify the approximation discussed in the previous section of neglecting  NP 
contributions to $e^+e^- \to$ hadrons. 
Finally, the two Majorana dark fermions are characterised by  a certain mass splitting $\Delta m_\chi$ and,  in particular, the lightest one 
may also play the role of a dark matter particle.
This model provides all the conditions required to shift the luminosity estimates based 
on measurements of Bhabha events and to generate additional di-muon events.
Since the most worrisome discrepancy among the various datasets is represented 
by the low values of $\shad$ reported by the three KLOE analyses,  
we fix the $V$ mass close to the KLOE CoM energy. More precisely,
since for KLOE08/12 $\sqrt{s} = 1.020$ GeV while for KLOE10 $\sqrt{s} = 1.0$ GeV, 
we fix $M_V=1.001$ GeV 
so that  the KLOE10 measurement is also affected by NP events.

In order to avoid bounds from light resonances  searches, the DP main decay channel must consist of multibody final states, including a certain amount of missing energy.
With $m_{\chi_1}+m_{\chi_2} < M_V $,   
$V$ decays proceed mainly via the chain $V\to \chi_1\chi_2 \to \chi_1\chi_1 e^+e^-(\mu^+\mu^-)$, with BR$(V\to \chi_1\chi_2)\sim100\%$ and BR$(V\to e^+e^-,\, \mu^+\mu^-)\propto \varepsilon^2$.
In particular, to ensure that the $e^+e^-$ and $\mu^+\mu^-$ events from $V$ decays
will carry away sufficient energy to populate the datasets 
after the experimental cuts, we choose $m_{\chi_2} \sim 0.97\, M_V \gg m_{\chi_1} \sim\,3$ MeV (the dark matter mass does not play a critical role as long as $m_{\chi_1}\lesssim\mathcal{O}$(10) MeV).

This is possible by assuming that a new boson produced resonantly around the KLOE center-of-mass (CoM) energy decays promptly yielding $e^+e^-$ and $\mu^+\mu^-$ pairs in the final state. 
This can give rise to three different effects:
\begin{enumerate}
\item the additional $e^+e^-$ events will affect the KLOE luminosity determination based on measurements of the Bhabha cross section, and in turn the inferred value of $\sigma_{\rm had}$;
\item  the additional $\mu^+\mu^-$ events will affect the determination of $\sigma_{\rm had}$ via the (luminosity independent) measurement of the ratio of $\pi^+\pi^-\gamma$ versus  $\mu^+\mu^-\gamma$ events;
\item  loops involving the new boson would give a direct contribution 
to the predicted value of $a_\mu$.
\end{enumerate}
All these  effects were  discussed in detail in  Ref.~\cite{Darme:2021huc}, where 
it was concluded  that  a new gauge boson $V$ of  mass 
close to the mass of the $\phi$ meson $M_V \sim M_\phi \simeq 1.020\,$GeV  
was able  to release the tensions between the  KLOE and  BaBar results for $\shad$, 
the data-driven and   lattice determinations of $\amhvp$, and the measured 
values of $a_\mu$ with the theoretical prediction, 
without conflicting  with other phenomenological constraints.
Yet, in  Ref.~\cite{Darme:2021huc} a complete agreement 
among all the datasets could not be reached, and this was essentially 
due to the fact 
that one of the three KLOE measurements of $\shad$ (commonly referred as KLOE10~\cite{KLOE:2010qei}) was performed at a CoM energy
20 MeV below the $V$ resonance, thus remaining unaffected by the NP.

While we adopt the same theoretical model of Ref.~\cite{Darme:2021huc},  here we shift downwards by a few MeV the location of the $V$ resonance. With respect to the analysis in  Ref.~\cite{Darme:2021huc} such a small shift leaves the NP effects on the BaBar dataset essentially unmodified, because it operated at a  CoM energy much larger than $M_V$ ($\sqrt{s}=10.6$\,GeV). For KLOE08 and KLOE12, that collected data at the $\phi$ resonance ($\sqrt{s}=1.02\,$GeV), the NP effects  are somewhat reduced  but, due to radiative return on the nearby $V$ resonance, are still significant. 
Most importantly, now the KLOE10 luminosity measurement get also affected by additional Bhabha events of NP origin, resolving the tension observed in Ref.~\cite{Darme:2021huc} between the KLOE10 and KLOE08/KLOE12  determinations of $\shad$.

\subsection{Window anomaly and GeV-scale new physics}

 \begin{table}[t]
	\centering
	\begin{tabular}{l |c|c|c|c}
	 & $a_\text{SD}^\text{HVP,\,NP}$ & $a_\text{W}^\text{HVP,\,NP}$ & $a_\text{LD}^\text{HVP,\,NP} $ & $a_\text{total}^\text{HVP,\,SM}$\\
	 \hline
	 \hline
	KLOE08 & $0.03$ & $0.31$ & $0.69$ & $368.7$ \\
 	KLOE10 & $0.65$ & $6.66$ & $15.07$ & $366.0$ \\
  	KLOE12 & $0.06$ & $0.63$ & $1.43$ & $366.6$ \\
    BaBar & \!\!\!$0.59$\!\!\! & $6.67$ & $15.68$ & $376.7$\\
    	CMD-3 & $ - $ & $ - $ & $ -  $ & $383.7$ \\
	\hline
	\end{tabular}
	\caption{
 Indirect new physics contribution to $\amhvp$ for the $\pi\pi$ channel 
 in  the short, intermediate and long distance windows for KLOE08~\cite{KLOE:2008fmq}, KLOE10~\cite{KLOE:2010qei}, KLOE12~\cite{KLOE:2012anl} and BaBar~\cite{BaBar:2009wpw,BaBar:2012bdw}, 
in units of $10^{-10} (\varepsilon/0.0125)^2$
and in the range $\sqrt{s'}=[0.6-0.9]$ GeV. 
 The last column shows the total $\pi\pi$ channel SM contribution in units of $10^{-10}$, computed in the relevant $\sqrt{s'}$ range from the data publicly available in the literature. The theoretical errors are not shown (see the discussion in the main text).
 }
	\label{tab:shift}
\end{table} 
Once the masses of the NP particles have been fixed, 
 we compute the absolute shift for KLOE08, KLOE10, KLOE12 and BaBar
as the function of the coupling $\varepsilon$. Table~\ref{tab:shift} shows the shift due to DP related events for the different experiments in the energy region $\sqrt{s'}~\in~[0.6,\,0.9]$\,GeV (in which all the experiments have provided the relevant information)
and for $\varepsilon = 0.0125$.
It can be seen that the indirect NP contribution in the short distance window is negligible with respect to the ones in the intermediate  and long distance windows.
This is because the corresponding weight function strongly suppresses the SD
NP contribution. 

\begin{figure}[t!]
    \centering
   \includegraphics[width=0.49\textwidth]{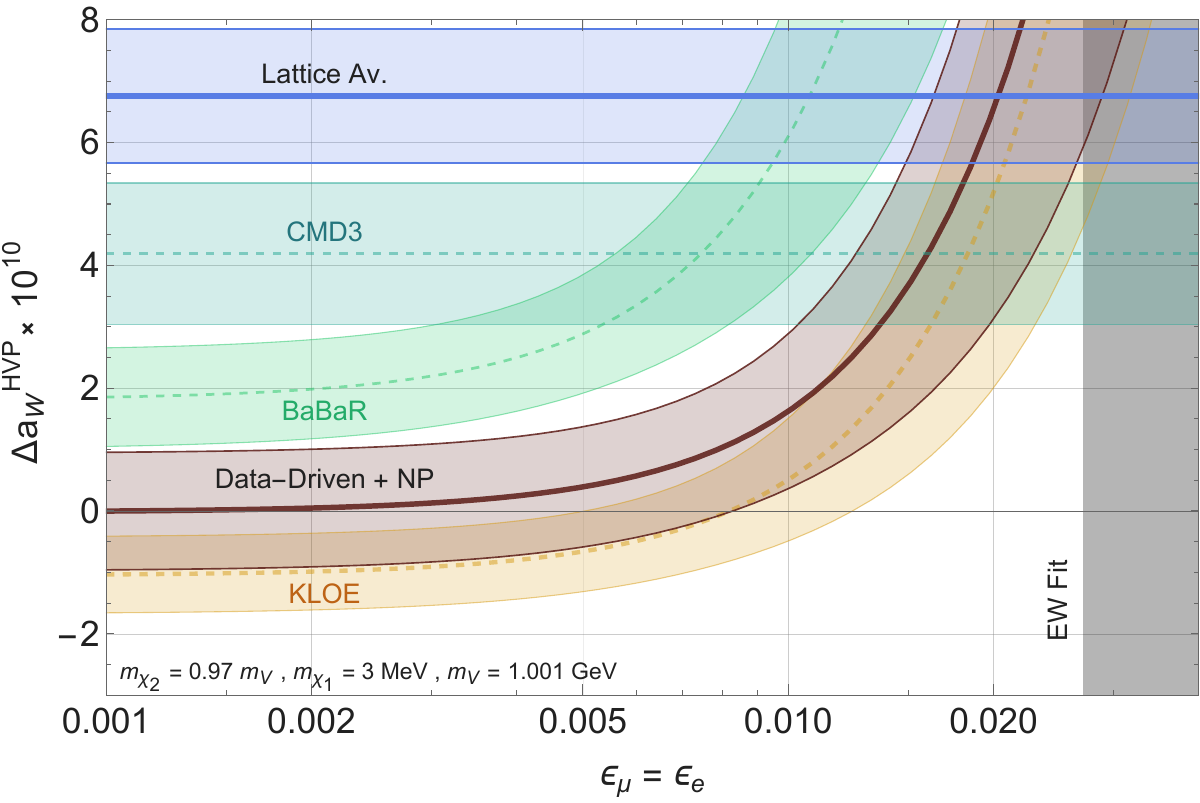}
    \caption{
    Theoretical estimate  for the shift in $a_\mu$ in the intermediate window compared to the data-driven result for KLOE (orange region), BaBaR (green region), CMD-3 (aquamarine region) and for the full data-driven combination (dark red region). All the bands show the $1\sigma$ regions. $\Delta \amIW =\amIW (\varepsilon)-\amIW(0)$ in the vertical axis is the NP contribution with respect to the SM data-driven estimate. The lattice result from Ref.~\cite{WittingMoriond2023} is shown in  blue. For reference we also depict in dark grey the excluded region from LEP 
for the case of a kinetically-mixed dark photon model.
  }
    \label{fig:final_plot}
\end{figure}

Fig.~\ref{fig:final_plot} shows the evolution of the shift in the 
contributions to $\amhvp$ from the intermediate window, resulting from 
our NP model, as function of the parameter $\varepsilon$. 
The dark region at $\varepsilon \geq 0.027 $ is excluded by the limit from electroweak precision observables  derived  in Refs.~\cite{Hook:2010tw,Curtin:2014cca} for the case in which $V$ is a kinetically-mixed dark photon. Note that in any case the direct loop contribution to $a_\mu$ significantly overshoots the experimental value.
We did not include in our estimates  possible effects from background subtraction and 
corrections to the efficiency, as they cannot be estimated in a reliable way. 
For this reasons we believe that the  NP effects are likely underestimated.

Several comments are in order. 
In first place,  the 
shifts due to  NP are estimated  only for  the energy range $\sqrt{s'}~=~[0.6,\,0.9]$ GeV, which corresponds to about half of the HVP contribution in the 
intermediate window.\footnote{We do not include the results on 
 $e^+ e^- \to K K$  provided by the 
SND~\cite{Achasov:2000am} and CMD-2~\cite{CMD-2:2008fsu} collaborations.
Since  $e^+ e^- \to K K$  is dominated by the $\phi$ resonance peak, similarly to KLOE08 
also these results will likely be  shifted due to their calibration of the luminosity via Bhabha scattering events. However, we can expect that the overall effects of these additional  measurements 
will be small,  due to both the large experimental errors and the small contribution from $KK$ final states to the total HVP.}
For $\sqrt{s'}$ above the GeV no NP effects are expected, 
since $e^+e^-$ ($\mu^+\mu^-$) events with such an invariant mass 
cannot originate from the $V$ resonance with $M_V \sim 1\,$GeV. 
However, the range $\sqrt{s'} < 0.6$ GeV will also contribute to the overall shift. In order to account for  (1) the missing data relative to the $\sqrt{s'} <0.6\,$GeV range, (2) the potential NP effects on background subtractions and on the efficiency calibration procedure,
and (3) the effect of experimental smearing described in Appendix~\ref{sec:app1}, we  include a $50 \%$ theoretical error on the overall size of the shifts generated by the indirect NP effects.

To estimate the  window contribution   $\amIW$ 
from the CMD-3 data in the energy range $\sqrt{s}\in[0.6,\,0.9]\,$GeV   
we have used the values of the pion form factor $|F_\pi|^2$ given in the ancillary files  of Ref.~\cite{CMD-3:2023alj} to compute the two-pion cross section, from which 
the  $\amIW$ contribution is derived. We obtain:
\begin{equation}
\label{eq:CMD3}
        a_{\rm W}^{\rm HVP, CMD-3}\Big|_{\sqrt{s}/{\rm GeV}\in[0.6,0.9]}= 114.5 (1.2) \cdot 10^{-10}\,.
\end{equation}
 Since in our scenario the evaluation of  $\amhvp$ from CMD-3 data does not receive   
 NP contributions ($a_{\rm W}^{\rm HVP,  NP}= 0$)  the value 
in \eqn{eq:CMD3}  remains constant across  all values of $\varepsilon$,
see Fig.~\ref{fig:final_plot}. 

An important improvement of the present study with respect 
to Ref.~\cite{Darme:2021huc} 
is that the residual internal discrepancy within the KLOE experimental data-sets that was formerly observed in correspondence with the largest values of $\varepsilon$ is now resolved. 
This is due to the fact that  in Ref.~\cite{Darme:2021huc} the $V$ mass was fixed at the value $M_V \simeq M_\phi$, so that the NP was affecting  KLOE08 and KLOE12, but not the KLOE10 data, that were taken at a CoM energy  20\,MeV  below the $\phi$ resonance.  
Once ISR effects are included, the slightly lower value $M_V \simeq M_\phi - 17\,$MeV  adopted in this paper is enough to mitigate this issue, since now  all the three KLOE datasets include $\varepsilon$-dependent NP contributions.

Finally, our estimate of the $a_{\rm W}^{\rm HVP}$ for CMD-3 in their full range $0.327\, \leq\, \sqrt{s}/\,{\rm GeV}\leq 1.199$ is:
\begin{equation}
\label{eq:CMD3full}
        a_{\rm W}^{\rm HVP, CMD-3}\Big|_{\sqrt{s}\leq1.2{\rm GeV}}= 139.4 (1.6) \cdot 10^{-10}\,.
\end{equation}

\subsection{Internal discrepancies of the $\shad$ datasets }
\label{sec:discrepancies}

The measurements of $\shad$ performed with the energy scanning
method is  not  affected by the NP
when the data points are taken at $\sqrt{s}< M_V$.
This is the case for the data points used by the 
CMD-3 collaboration to compare their results with the ones 
of the other experiments, which fall in the 
interval $\sqrt{s} \in [0.60,0.88]\,$GeV~\cite{CMD-3:2023alj}.
Note that the CMD-3 result is a couple of $\sigma$ above BaBar,  several $\sigma$ above the combined KLOE result, and quite 
close to the lattice estimate, in nice agreement with what is expected in our scenario.
However, for the same reason also the  
CMD-2~\cite{Akhmetshin:2003zn,Akhmetshin:2006wh,Akhmetshin:2006bx} and 
SND~\cite{Achasov:2006vp,SND:2020nwa} results are not  modified. 
This implies that the NP scenarios discussed above  cannot  mitigate  
the discrepancy between CMD-3 and CMD-2/SND data-sets.

Altogether, nine experimental results are included in our fit. While the $\sqrt{s}$ regions probed by each analysis differ, they all overlap in the $\sqrt{s} \in [0.60,0.88]\,$ interval. As a first measure of the (dis)-agreement between the different determinations of $\shad$, we estimate the p-value of $\amhvp$  for the corresponding 
data-sets, within the overlapping  CoM energy range.\footnote{Note that the KLOE08 and KLOE12 measurements are partially correlated~\cite{KLOE:2012anl}.} We find that in the SM-only hypothesis, the data-sets present a discrepancy at the $\sim 4.5 \sigma$ level. Note that this p-value is not linked directly to the window anomaly, and simply represents a test of the self-consistency of the different data-sets under the SM-only hypothesis.

We have also estimated the likelihood including the consistency in the intermediate window  of the averaged data-driven result $\amIW |_{\textrm{data}}$ with the lattice estimate $\amIW |_{\textrm{lat}}$, as well as the consistency between the FNAL  experimental result $a_\mu^{\textrm{exp}}$ and the prediction based on the BMW lattice result.
For this latter case there is substantial agreement with the SM-only case, however, our NP scenario also implies 
a direct (loop-induced)  contribution to $a_\mu$, that  should be removed from $a_\mu^{\textrm{exp}}$
to obtain the SM-only prediction based on the BMW's HVP result.

\begin{figure}[t!]
    \centering
   \includegraphics[width=0.49\textwidth]{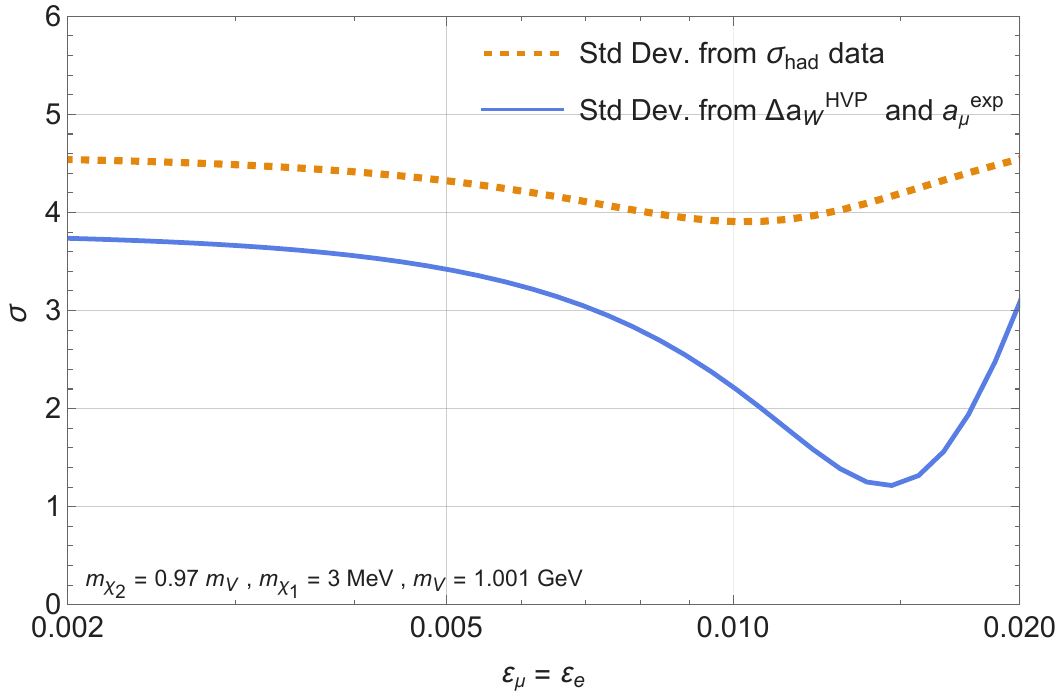}
    \caption{
    Standard deviation of the global fits to the internal consistency of the data-driven data-set, and to the window anomaly as a function of $\varepsilon$ for $m_{\chi_1} = 3 $ MeV, $m_{\chi_2} = 0.97 \,m_V \,$, $m_V=1.001$ GeV and $\alpha_D=0.5$.  
  }
    \label{fig:chi2lot}
\end{figure}

\begin{figure}[t!!]
    \centering
    \includegraphics[width=0.49\textwidth]{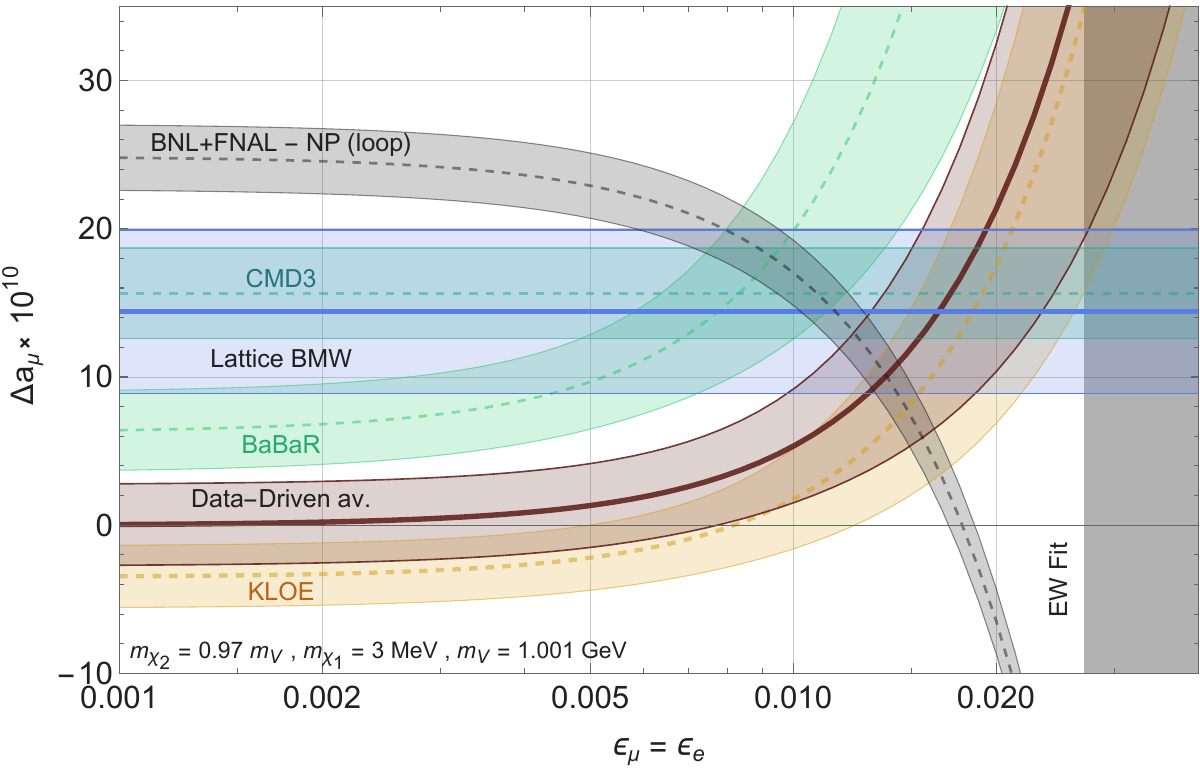}
    \caption{
Shifts in the prediction of the data-driven results from KLOE (orange band), BaBar (green band), CMD-3 (aquamarine band) and  from the average of the $e^+e^-$ data (brown band), 
as a function of $\varepsilon$  for $m_{\chi_1} = 3 $ MeV, $m_{\chi_2} = 0.97 \,m_V \,$, $m_V=1.001$ GeV and $\alpha_D=0.5$. 
The gray band corresponds to the 
experimental world average \eqn{eq:amuexp} with the direct NP contribution to $a_\mu$ from NP loops subtracted out.
For reference we also depict in dark grey the excluded region from LEP 
for the case of a kinetically-mixed dark photon.}
    \label{fig:new_plot}
\end{figure}

We show in Fig.~\ref{fig:chi2lot} the result of both global fits, expressing for convenience the p-value in term of standard deviations  for our benchmark parameter point. The window discrepancy  corresponds to the blue curve, and we see that for values of the coupling above $\varepsilon\sim 0.012$ it decreases well below $2\sigma$. In contrast,  the inner tension within the experimental data-sets, that  correspond to the orange dashed curve, is only very mildly mitigated  
regardless the value of  $\varepsilon$.
Finally, in Fig.~\ref{fig:new_plot} we illustrate the  impact of the direct and indirect  NP effects on the overall scenario  of the $a_\mu$ related discrepancies. We see that a remarkable improvement in the agreement among the various theoretical and experimental determinations can be obtained for $\varepsilon \sim O(1\%)$.

The fact that our  mechanism  can solve the discrepancies between lattice  and data-driven 
results both for the total and intermediate windows HVP contributions
without invoking modifications of $\shad$ above 1~GeV, is consistent with the literature on this topic~\cite{Colangelo:2020lcg} in that it relies on a local change within the $\sqrt{s} \in [0.60,0.9]\,$ region, and not just on an uniform shift in the low-energy $\pi\pi$ region.
This possibility has been also confirmed by the recent dedicated analysis presented in   Ref.~\cite{Davier:2023cyp}, where it is shown that the results of the dispersive method can be reconciled with those of the lattice approach by modifying the hadronic experimental spectrum in the $\rho$ region.
On the other hand,  our model 
does not succeed in reconciling all the discrepancies between the various $\shad$ data-sets and, as we have seen,  
 the improvement with respect to the SM-only hypothesis remains  marginal for all values of $\varepsilon$.
This issue has two facets. The first one is that the NP shift is larger for BaBar (and KLOE10) than for KLOE08/12.
However, without NP corrections BaBar results are already above KLOE.
This situation can be improved by adding another dark photon sitting at a somehow larger mass, corresponding to the KLOE08/12 CoM energy. We have estimated that with this setup the overall tension in the 
$\shad$ data-set can be reduced to around $2.6\sigma$, see Appendix~\ref{sec:app2}.
The fact that the overall agreement cannot be further improved is because of the additional large discrepancies that are present between the results of experiments that rely on the energy-scanning method, and in particular CMD-3 vs. CMD-2/SND. Indeed these experiments would be 
affected in similar ways in any NP scenario,  we thus believe that this second facet points towards the possibility of experimental problems.

 \section{Conclusions}
 \label{sec:conclusions}

In this work we have studied  the possibility of exploiting  NP processes to solve  
the discrepancy between the dispersive and lattice estimates of $\amhvp$ in the 
so-called \emph{intermediate} window, that has persisted throughout different lattice determinations, 
and has by now reached a worrisome level of significance.
 The key NP ingredient is a GeV-scale new boson which is produced 
 in $e^+ e^-$ collisions, and that decays promptly  into 
 $e^+ e^-, \mu^+ \mu^-$ + missing energy.  
   The amount of  missing energy in the decay final state must be sufficient  to avoid current constraints,  and is provided by light dark sector fermions produced in the decay chain. This NP scenario can 
 affect  the prediction for $a_\mu$  {\it directly} via new loop contributions, 
 but most interestingly it can {\it indirectly} affect the determination 
 from experimental data of $\shad$,  which is used to derive the  theoretical prediction for  $a_\mu$  by means of the dispersive  approach. 
 We have argued that our simple phenomenological model    
 can sizeably reduce the `window' tension, to a significance  at   
 the $\ 1 \sigma$ level. We believe that this result can pave the way
 to build UV complete models able to explain the  $(g-2)_\mu$ anomalies.
Additionally, with respect to Ref.~\cite{Darme:2021huc}, the present work improves the technical aspects of the estimation of the indirect effects of GeV-scale NP on the data-driven $\amhvp$ in several important directions. In particular, (i) we made a quantitative estimate of the  NP effects on $\amhvp$ for the BaBar dataset, (ii) we included NLO effects which modified strongly the NP shifts on the KLOE dataset and (iii) we performed a global fit of the data-driven data including all main existing analysis.  We also made a quantitative estimate of the intermediate window contribution to $\amhvp$ inferred from the recently published CMD-3 data~\cite{CMD-3:2023alj} which, in agreement with what could have been expected in our scenario, does not show any particular tension with the lattice result. 

 From this study a certain number of conclusions can be drawn. First, indirect  effects of new GeV-scale particles on the measurements used to estimate $\amhvp$ via the dispersive method can have important consequences, and can be significantly more ubiquitous than what was anticipated in Ref.~\cite{Darme:2021huc}. In particular, NP effects can bias the determination of $\shad$ 
 via the efficiency calibration and background removal processes used by the experimental collaborations. Second, in our scenario datapoints collected at CoM energies below 1~GeV are not affected by NP effects.
This nicely explains the qualitative agreement of the recent CMD-3 result~\cite{CMD-3:2023alj} with the lattice, as well as the significant discordance 
with the results of other experiments operating at $\sqrt{s} \gtrsim M_V $, that have measured  
 $\shad$ in the same energy range by exploiting the radiative method.
On the other hand, 
for the same reason also the CMD-2~\cite{Akhmetshin:2003zn,Akhmetshin:2006wh,Akhmetshin:2006bx} and 
SND~\cite{Achasov:2006vp,SND:2020nwa} results are not  modified by this NP, and thus  the discrepancy 
with the CMD-3 result endures. 
Third, for experiments running at CoM energies at or above 1~GeV, NP effects in the determination of $\amhvp$  are negligible 
for the short-distance window, while they are   sizeable in the intermediate (and likely also in the long-distance) window. 
Fourth, the key ingredient to evade existing limits from searches of  light, weakly
coupled new particles, is the presence of a significant amount of missing energy 
associated with their visible decay products. 
At the same time, as long as $e^+e^-$ ($\mu^+\mu^-$) pairs are present  in the  
 (multibody) final state, due to the not-too-tight cuts 
typically set by the experimental collaborations in measuring $\shad$,  
the missing energy requirement  does not preclude  significant NP effects on 
the determination of $\amhvp$.  

\vspace{0.5cm}
\noindent \textbf{Note added:} After the first submission of this paper to the {\tt arXiv} in December 2022  
several new  results related to the $(g-2)_\mu$ anomalies, and in particular to the  photon HVP appeared. 
In January 2023 the results of the Fermilab Lattice, HPQCD, and MILC~\cite{Bazavov:2023has} and RBC/UKQCD RBC/UKQCD~\cite{Blum:2023qou} appeared,  giving further support 
to the reliability of the lattice estimates of the HVP. Most importantly, in February 2023 the  CMD-3 collaboration 
announced their new result on the $e^+e^-\to \pi^+\pi^-$ cross section~\cite{CMD-3:2023alj}, which is in dramatic 
tension with the previous KLOE (and, to a lesser extent, also BaBar) result, while it is in broad agreement with 
the lattice. This nicely fits  within our scenario and supports its plausibility. The original article has thus been 
expanded  to account for  these important results. 
Additionally, towards the completion of this revised version other  
 new results appeared which also strengthen the conclusions of this work, and that have been integrated 
 in the text as well. First, a new experimental measurement of the 
 muon anomalous magnetic moment by the FNAL Muon $(g-2)$ experiment~\cite{Muong-2:2023cdq} confirmed the first 
 measurement by the same collaboration, yielding the present world average in~\eqn{eq:amuexp}. Second, 
as was pointed out in footnote \ref{foot:BaBar},
a study of NNLO effects by the BaBar collaboration~\cite{BaBar:2023xiy} remarked that it may 
have a significant effect for some of the 
KLOE analyses, and this might reduce the tensions between the 
 $\shad$  data-sets that  our model cannot explain satisfactorily. Finally, as was mentioned in Section~\ref{sec:discrepancies},
 the new study addressing  the discrepancies between the lattice QCD and data-driven  
results presented in Ref.~\cite{Davier:2023cyp} has shown that  the two approaches can be brought into agreement by modifying the $\rho$ peak 
in the experimental spectrum. This is precisely what the indirect effects generated by our model do.

\bigskip

\section*{Acknowledgments}

We thank Bogdan Malaescu for providing us with information 
on the BaBar measurement of $\sigma_{\pi\pi\gamma_{\rm ISR}}$.
This work has received support from the INFN ``Iniziativa Specifica" Theoretical Astroparticle Physics (TAsP-LNF) and from 
the European Union’s Horizon 2020 research and innovation programme under the Marie Sklodowska-Curie grant agreement No. 101028626.
The work of G.G.d.C. was supported by the Frascati National Laboratories (LNF) through a Cabibbo Fellowship, call 2019. 
The work of E.N. was supported by the Estonian Research Council grant PRG1884.

\appendix
\section{Experimental cuts and smearing}
\label{sec:app1}

We  present the main ingredients of the  recasting of the $e^+ e^- \to e^+ e^- (\gamma^{\rm ISR})$ 
and  $e^+ e^- \to \mu^+ \mu^- (\gamma^{\rm ISR})$ experimental analysis for our NP events.

We have simulated all NP events in the~\amc~framework, then smeared the momenta of the final states particles to reflect the experimental precision. For KLOE, we use~\cite{KLOE:2012anl}:
\begin{align}
    \sigma_{p_T} &= 0.4 \% \times p_T \nonumber \\
    \sigma_{E} &= 5.7 \% \times \sqrt{E \ \textrm{(GeV)}} \, ,
\end{align}
with   $\sigma_{p_T}$  corresponding to the typical precision on charged momenta tracks and  
$\sigma_{E}$ to the precision  on the photon energy reconstruction (the polar angular precision is around $1^\circ$ and is further included to obtain the energy of charged tracks).\footnote{We have checked that with this smearing parameter we could reproduce with good accuracy the SM muon track mass distribution given in~\cite{KLOE:2012anl}.}  For BaBar, we use only the energy smearing~\cite{Ruland:2009zz}:
\begin{align}
    \sigma_{E} &= (2.3 \% \times E^{3/4}) \oplus (1.35\% \times E) \ \  \ \textrm{($E$ in GeV)} \ .
\end{align}

The  smearing effects typically broaden the $\sqrt{s^\prime}$ range where NP effects are relevant, and increase the  selection rates for NP  events, since they tend to ``hide'' the associated missing energy.
For all the experimental processes listed below, we have  applied the selection cuts in two steps: first a ``broad'' selection with weaker cuts is applied at the MC truth level, then an exact selection with the experimental cuts is applied after momenta smearing for the charged and photon tracks. We have applied the same procedure on both NP and SM events, and then we have used the ratio of efficiencies to derive the final shifts on $\amhvp$ and $\amIW$.

\vspace{0.2cm}

Below we list for convenience  the selection cuts for each analysis as reported by the experimental collaborations.
\paragraph{KLOE08 and KLOE10} Both analysis~\cite{KLOE:2008fmq,KLOE:2010qei} relied on Bhabha scattering to calibrate the luminosity. The experimental cuts are given by~\cite{KLOE:2006itf}: $|\cos \theta_{e^\pm}| < 0.57$, $E_{e^\pm} \in [0.3,0.8] \, \rm GeV$, $|\vec{p}_{e^\pm}| \geq 0.4$ GeV and a cut is applied on the polar angle acollinearity of the $e^+$ and $e^-$ charged tracks: $\zeta \,\equiv\, |\theta_{e^+} + \theta_{e^-} - 180^\circ | < 9^\circ$. 

\paragraph{KLOE12} The KLOE12~\cite{KLOE:2012anl} analysis used kinematic cuts on the two muons polar angles  $|\cos \theta_\mu \,|<0.64$, momenta $p^T_{\mu} \geq 160 \, \rm MeV$ or $|p^z_{ \mu} |\geq 90$ MeV, and a cut on the polar angle of the missing photon (as reconstructed from the observed muons momenta) $|\cos \theta_\gamma \, | > \cos( 15^\circ)$. Finally, the reconstructed track mass  of the $\mu^+ \mu^-$ system, as defined in Eq.~\eqref{eq:mtr} (see e.g.~\cite{KLOE:2010qei}), must satisfy $m_{tr} \in [80,115]$ MeV. This last cut is by far the most stringent one because $m_{tr}$ is very sensitive to the presence of missing energy. As an example, varying the smearing of $\sigma_{p_T}$ by a factor of two leads to a variation of the NP cut efficiency by around $40 \% -50 \%$,  depending on the parameter point.

Note that for all the three KLOE analyses, the final cross-section is obtained by including soft initial state radiation  
according to the analytic expressions 
given in ref.~\cite{Nicrosini:1986sm}. 

\paragraph{BaBar} In the last analysis of the BaBar collaboration~\cite{Lees:2013rw} the following selection cuts were applied: polar angles of charged tracks in the laboratory frame are required to be in the range $\theta_{\mu^\pm} \in [0.35 , 2.4] \textrm{ rad} $, and   $\theta_{\gamma} \in [0.45 , 2.45] \textrm{ rad} $ for photon tracks. The photon energy in the CoM frame must satisfy $E_\gamma^* > 3$ Gev, and the charged track momenta $|\vec{p}_{\mu^\pm}| > 1 $ GeV. Additionally,  a preselection cut requiring that the ISR photon lies within $0.3$ rad of the missing momentum of the charged tracks in the laboratory frame was also applied.

\section{New physics versus experimental data-sets tensions}
\label{sec:app2}

\begin{figure}[t!!]
\vspace{-0.5cm}
    \centering
        \includegraphics[width=0.49\textwidth]{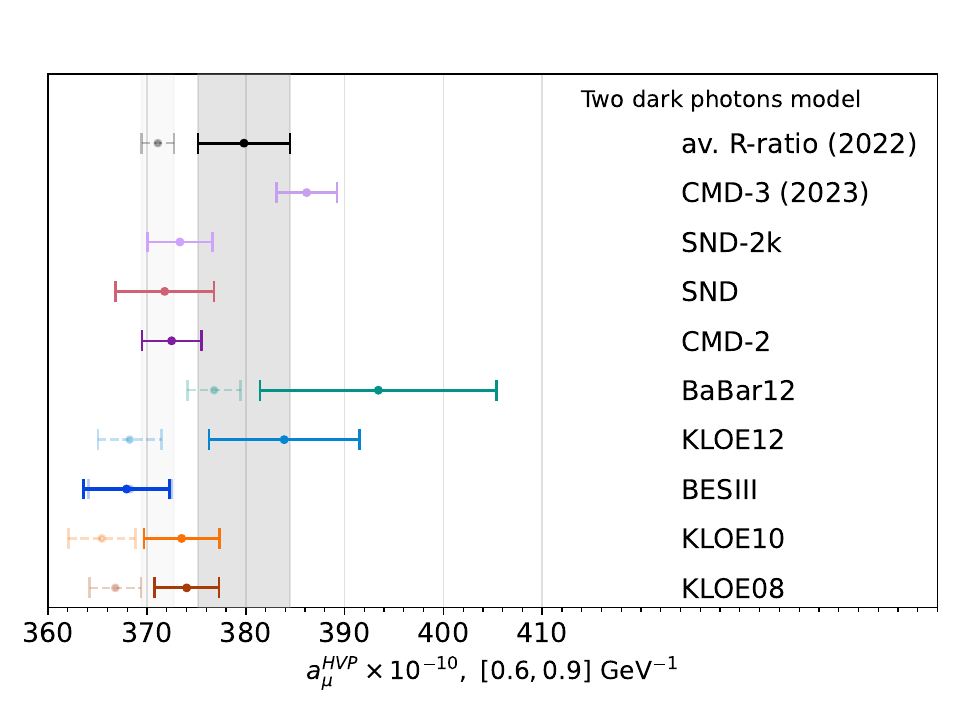}
    \caption{
Shifts in the prediction for the data-driven results for all the experimental data-sets included in our analysis.
The shifts are given in terms of $\amhvp$ values in the  $\sqrt{s} \in [0.60,0.9]\,$ region, 
both for SM-only hypothesis (dashed lines)  and for  the best-fit model with two dark photons,  with $m_{\chi_1} = 3 $ MeV, $m_{\chi_2} = 0.97 \,m_V \,$, $m_{V1}=1.001$ GeV and  $m_{V2}=1.0175$ GeV, $\alpha_D=0.5$ and $\varepsilon_1 = 0.012 , \varepsilon_2 = 0.008$ (solid lines). The gray band depicts the average of the $e^+e^-$ data.}
    \label{fig:ShiftData2DP}
\end{figure}

The main limitation of the NP model we used so far is the fact that  large changes in the luminosity determination cannot be obtain  for KLOE08, KLOE10 and KLOE12 simultaneously. 

Thus it is interesting to explore if this limitation can be circumvented by extending our model with on additional dark photon, $V_2$.  This attempt should  be understood as a practical way to assess the best results achievable with this class of models.
In Fig.~\ref{fig:ShiftData2DP}  we plot the experimental shifts for the nine experimental results used in this work, choosing $m_{V1}=1.001$ GeV, $\varepsilon_1 = 0.012$ and  $m_{V2}=1.0175$ GeV,  $\varepsilon_2 = 0.008$ for the first and second dark photons respectively. The error bars reflect both the experimental error as well as an educated guess of the theoretical errors associated with additional NP effects not included in the analysis, that could for example impact 
the experimental calibration of the efficiencies and the background subtraction procedure.
Even with this ad hoc setup, the tension in the data-set 
remains at the $2.6\sigma$ level. This suggests that it is unlikely that all the discrepancies among the different data-sets (as, for example, between the CMD-3 and CMD-2 results) 
could be accounted for by some NP.

\vfill

\bibliographystyle{apsrev4-1}

\bibliography{bibliography}

\end{document}